\documentclass[preprint,preprintnumbers,amsmath,amssymb]{revtex4}
\usepackage{graphicx}
\usepackage{dcolumn}
\usepackage{bm}
\usepackage{epsfig}

\newcommand{\geqnew}{\stackrel{>}{\!\ _{\sim}}}

\newcommand{\prefactor}{\frac{1}{\sqrt{2\pi \hbar}}}
\begin{document}
\title{Visualizing classical and quantum probability densities for
momentum using variations on familiar one-dimensional potentials}

\author{R. W. Robinett} \email{rick@phys.psu.edu}
\affiliation{%
Department of Physics\\
The Pennsylvania State University\\
University Park, PA 16802 USA \\
}%

\date{\today}

\begin{abstract}

After briefly reviewing the definitions of classical probability
densities for position, $P_{CL}(x)$, and for momentum, $P_{CL}(p)$,
we present several  examples of  classical mechanical potential
systems, mostly variations on such familiar cases as the infinite well
and the uniformly accelerated particle  for which the classical 
distributions can be easily derived and
visualized.  
We focus especially on a simple potential which interpolates between the
symmetric linear potential, $V(x) = F|x|$, and the infinite well, which
can illustrate, in a mathematically straightforward way, 
 how the divergent, $\delta$-function classical probability
density for momentum for the infinite well can be easily seen to arise. 
Such examples can
help students understand the quantum mechanical momentum-space wavefunction
(and its corresponding probability density) in much the same way that other
semi-classical techniques, such as the WKB approximation, can be used to
visualize position-space wavefunctions.

\end{abstract}

\maketitle

\begin{flushleft}
{\large {\bf 1.~Introduction}}
\end{flushleft}
\vskip 0.2cm

Since the introduction of quantum mechanics early in the last century,
practitioners and students of the subject alike have used semi-classical
methods to calculate, understand, and visualize properties of 
position-space solutions of the time-independent Schr\"odinger equation.
WKB-type  solutions \cite{park}, for example, provide an attractive 
and mathematically
understandable approximation which makes manifest the typical correlations
between the magnitude of a stationary state wavefunction and the corresponding
'wiggliness'. At a somewhat less sophisticated level, many  textbooks 
\cite{liboff}, \cite{robinett_book},
make direct comparisons between the quantum mechanical probability 
density, defined by $P_{QM}(x) = |\psi(x)|^2$, with intuitively derived 
classical counterparts, $P_{CL}(x)$, which are often motivated by very 
simple '{\it how much time is spent in a given interval}' arguments.

In comparison, there are far fewer discussions in the pedagogical literature
or in standardly used textbooks 
giving examples of the corresponding semi-classical connections to 
momentum-space solutions, $\phi(p)$. Solutions to the Schr\"odinger
equation obtained directly from a momentum-space formulation (such as
for the uniformly accelerated particle) are sometimes presented, but
most often the information encoded in $\phi(p)$ is simply noted to be
related to that in the position-space wavefunction by a rather formal
operation, namely the Fourier transform, via
\begin{equation}
\phi(p) = \prefactor \int_{-\infty}^{+\infty} \, \psi(x)\, e^{ipx/\hbar}\,dx
\,.
\end{equation}
Correlating the information clearly evident in the magnitude/wiggliness
of a position-space, $\psi(x)$, with the similar information encoded in 
$\phi(p)$ is a non-trivial task,  even for experienced scientists, and
students first approaching the subject often can find no obvious connections
between the form of the quantum mechanical probability density for 
momentum, $|\phi(p)|^2$, and their intuition about classical mechanics.
A recent study \cite{qmvi} of the progression of student understanding 
of quantum mechanical concepts, starting with introductory modern physics 
courses, 
through advanced undergraduate quantum mechanics classes, to first year
graduate quantum theory courses,  has provided evidence that while student
understanding of the form of position-space wavefunctions does increase
through the undergraduate curriculum, their ability to extend these notions
to momentum-space ideas lags far behind. 

Finally, since the results of experiments from a wide variety of areas
in physics are increasingly presented in the form of momentum distributions
\cite{modern_1}, \cite{modern_2}, a background in understanding
a momentum-space approach to classical and quantum mechanics is becoming 
more important for students to acquire.

Motivated  by all of these factors, in this note we will discuss  
several simple model systems for which
one can rather easily calculate, analyze, and visualize 
 the classical probability densities for both position and momentum,
and to also compare them directly to their quantum mechanical
equivalents to visualize the correspondence principle limit in both $x$-
 and $p$-space. We first briefly review, in Sec.~2,  the basic definitions of
classical probability densities for one-dimensional bound state problems
and apply these methods to the simple problem of the 'bouncer', a particle
subject to a uniform gravitational force, but with a rigid, infinite
wall constraint due to a horizontal surface. 
As our main example, in Sec.~3, we provide an attractive interpolating case 
which shows how both the classical and quantum mechanical descriptions for 
the symmetric linear well (defined by $V(x) = F|x|$ and hence related to the
'bouncer' or uniform acceleration case) and the standard infinite
well potentials can be smoothly connected to one another, and how the
classical momentum distributions can be easily extracted.  we also show,
in some detail, how the magnitude/wiggliness of the position space quantum 
wavefunction is directly correlated to the form of its momentum-space 
$\phi(p)$ transform.
This last case also provides an intuitively clear and 
mathematically appropriate description of how the limiting case of the
divergent, $\delta$-function momentum-space classical probability densities
can arise for the infinite well case.

\begin{flushleft}
{\large {\bf 2.~Review of classical probability densities for
position and momentum: the 'bouncer' as an example
}}
\end{flushleft}
\vskip 0.2cm

In order to review the calculation and visualization of classical
probability densities for position, $P_{CL}(x)$, and momentum $P_{CL}(p)$,
we begin by illustrating,  in Fig.~1, a rather generic potential energy 
function, $V(x)$ versus $x$, which can  give rise to bound states
which can  be described either classically or quantum mechanically. A
fixed value of the energy $E$ is shown which then defines the classical
turning points, $a$ and $b$, defined by $V(a) = E = V(b)$: this implies
that $P_{CL}(x)$ will be non-vanishing only over the range $(a,b)$.
The corresponding range in allowed momentum values are given by 
conservation of energy using
\begin{equation}
p(x) = \pm \sqrt{2m(E-V(x))}
\qquad
\mbox{giving}
\qquad
p_m \equiv \sqrt{2m(E-V_{min})}
\end{equation}
so that the classical probability density, $P_{CL}(p)$, 
 will be defined over the range $(-p_m,+p_m)$. The variation of 
momentum values with position in the potential well is perhaps best 
shown in a phase-space $x-p$ plot, also included 
in Fig.~1, which simultaneously illustrates the allowed ranges in both variables.

We note that a particle at a given $x$ location will experience both
positive ($+p(x)$) and negative ($-p(x)$) values with equal probability
as it moves from left to right and then back from right to left. Thus,
even if the potential energy function is not symmetric, the classical
probability density for momentum, $P_{CL}(p)$, will necessarily satisfy
$P_{CL}(-p) = P_{CL}(+p)$. This feature is also echoed in the $x-p$ plot
in Fig.~1 where the phase-space curve need not be symmetric about any vertical
axis, but will necessarily be symmetric about the $p=0$ line.

We note from the figure that there may
well be more than one location where the particle has the same 
momentum value, given by $p = \sqrt{2m(E-V(x))}$, 
 such as at $x = c$ and $x = d$. We must
then take both contributions to $P_{CL}(p)$ into account in the formalism
which follows. One case in which this may not occur is if there is an
infinite wall type potential barrier from which particles simply 'bounce'
and rebound instead of slowing down on one side, coming instantaneously
to rest, and then reversing direction as shown in the case in Fig.~1. 

The classical motion in any such bound state system will be periodic with
period $T_{CL}$, corresponding to a complete set of left to right and
back to left (e.g., $a$ to $b$ and then back to $a$) motions. For some 
purposes, we only require the 'half-period' (the $a$ to $b$ time) 
given by $\tau = T_{CL}/2$
and this quantity can be easily calculated from the potential energy
function via
\begin{equation}
\tau = \int_{t_a}^{t_b}\,dt = \int_{a}^{b}\,\frac{dx}{v(x)}
= \sqrt{\frac{m}{2}}\int_{a}^{b} \frac{dx}{\sqrt{E-V(x)}}
\,.
\label{period_definition}
\end{equation}

A standard argument leading to the classical probability density for
position, $P_{CL}(x)$, begins by noting that the small probability of
finding the particle with position in the range $(x,x+dx)$ should be
proportional to the amount of time spent in that interval, $dt$. The
fraction of time spent in that small interval will be $dt/T_{CL}$ and
using the definition of probability density, we find that
\begin{equation}
P_{CL}(x)\,dx \equiv d\mbox{Prob}[x,x+dx] = \frac{dt}{T_{CL}}
\,.
\label{prob_definition_x}
\end{equation}
We then recall that (i) the particle will be in the same $dx$ bin on both its
'back' and 'forth' trips,  so the last term should really be doubled, and (ii)
the classical speed, $v(x) = |dx/dt|$, can be used to relate $dx$ directly to
$dt$,  giving the familiar result
\begin{equation}
P_{CL}(x) = \frac{d\mbox{Prob}[x,x+dx]}{dx} = 
2\left(\frac{dt/T_{CL}}{dx}\right)  = \frac{1}{\tau v(x)}
\label{classical_position}
\end{equation}
where $\tau = T_{CL}/2$ is the 'half-period'. We note that 
Eqn.~(\ref{period_definition}) ensures that this definition will give
a $P_{CL}(x)$ which is automatically  normalized to unity.

A similar expression for
the small probability that a measurement will find the particle with
momentum values in the range $(p, p+dp)$ is given by
\begin{equation}
P_{CL}(p)\,dp \equiv d\mbox{Prob}[p,p+dp] = \frac{dt}{T_{CL}}
\label{prob_definition_p}
\end{equation}
and we need the definition of the classical force,  $|F| = |dp/dt|$, 
to be able to write
\begin{equation}
P_{CL}(p) \equiv \frac{d\mbox{Prob}[p,p+dp]}{dp} = 
\frac{dt/T_{CL}}{dp} = \frac{1}{T_{CL}|F(p)|}
\label{classical_momentum}
\end{equation}
where we require the magnitude of the force, as a function of $p$. 
Depending on the nature of the potential, we may have to take two or more
contributions into account when evaluating Eqn.~(\ref{classical_momentum}),
such as with the $x =c,d$ points in Fig.~1 mentioned above.

As an example of a calculation and visualization of both
$P_{CL}(x)$ and $P_{CL}(p)$ for a simple potential, we consider the
classical 'bouncer', corresponding to a particle bouncing, without loss
of energy, under the influence of gravity, over a rigid horizontal
surface \cite{bouncer_6}. The corresponding potential energy function 
is given by
\begin{equation}
      V(z) = \left\{ \begin{array}{ll}
               +\infty & \mbox{for $z<0$} \\
               mgz & \mbox{for $z \geq 0$}
                                \end{array}
\right.
\,.
\label{bouncer_potential}
\end{equation}
If we imagine a point object with total energy $E$, the maximum height, $H$,
(which is one turning point,  in addition to  $z=0$  where it 'bounces'),
is given by $E = mgH$; the maximum momentum magnitude, $p_m$,  is given by
$E = p_m^2/2m$ or $p_m = \sqrt{2mE}$. The 'half-period', $\tau$, is simply
the time it takes to fall through the distance $H$, namely $\tau = \sqrt{2H/g}$
and $T_{CL} = 2\tau$. The position dependent speed is given  
$v(z) = \sqrt{2/m}\sqrt{E-mgz}$ so that the classical position probability
density is
\begin{equation}
P_{CL}(z) = 
\frac{1}{\tau v(z)} 
= \left[\frac{1}{\sqrt{2H/g}}\right]
\left[\frac{1}{\sqrt{(2/m)(E-mgz)}}\right]
= \frac{1}{2\sqrt{H(H-z)}}
\label{bouncer_prob_x}
\end{equation}
which has an (integrable) divergence at $z=H$ where the particle slows
down and reverses direction, but is finite at $z=0$ where the particle
simply 'bounces', reversing direction, but with no change in speed.
The corresponding classical probability density for momentum will be 
non-vanishing only over the range $(-p_m,+p_m)$ and is given by
\begin{equation}
P_{CL}(p) = \frac{1}{T_{CL}|F(p)|}
= \left[\frac{1}{2\sqrt{2H/g}}\right]\left[\frac{1}{mg}\right]
= \frac{1}{2\sqrt{2m(mgH)}} = \frac{1}{2\sqrt{2mE}} = \frac{1}{2p_m}
\label{bouncer_prob_p}
\end{equation}
corresponding to a 'flat' momentum distribution over the allowed range.
For this potential, due to the infinite wall' at $z=0$, the momentum 
only achieves a particular value {\it once} during each complete cycle
(and not at two separate locations, such as the $x=c,d$ case in Fig.~1),
so we only have one contribution to $|F(p)|$.  (The quantum mechanical
solutions, for both $\psi(x)$ and $\phi(p)$, and their comparisons to the 
classical results,  for a very similar problem are discussed, in details,
in the next section.)

The expressions in Eqns.~(\ref{bouncer_prob_x}) and (\ref{bouncer_prob_p})
for $P_{CL}(x)$ and $P_{CL}(p)$ can be visualized using a 'projection of
trajectory' technique \cite{robinett_ajp} which implements the definitions 
of the classical probability densities in Eqns.~(\ref{prob_definition_x})  and
(\ref{prob_definition_p})  involving 'time spent' arguments.
We can use a typical solution to the 'bouncer' problem of the form
\begin{equation}
z(t)  =  v_0 t - gt^2/2 
\qquad
\quad
\mbox{and}
\quad 
\qquad
v_z(t)  =  v_0 - gt
\end{equation}
where $v_0 = \sqrt{2gH}$ is the initial speed, 
corresponding to an object 'thrown upwards'; the motion then repeats itself
after each classical period. 
In Fig.~2, on the respective vertical axes ($z(t)$ and $p_z(t) = mv_z(t)$), we
indicate equal-sized bins ($dz$ and $dp$), project them horizontally
until they intersect the trajectory curves, then 'drop down' to the
$t$ axes. The small amounts of time, $dt$,  corresponding to each 'bin'
in either $z$ or $p_z$ are  then proportional to the probability in that
interval and are  shown (as a histogram) on the appropriate vertical
axes. The characteristic peaking near $z=H$ and the finite value at $z=0$
is evident for $P_{CL}(z)$, while the striking 'flat' momentum probability
density is easily visualized using this technique. A number ($N=1000$) 
of random 'measurements' of the particle's position and momentum over a single 
classical period are also shown alongside the 'binned' probabilities 
(as dots) and reflect the same phenomena.

\begin{flushleft}
{\large {\bf 3.~The infinite well as a limiting case
}}
\end{flushleft}
\vskip 0.2cm

The most familiar of all quantum mechanical bound state problems,
the infinite well, can also be used an example of classical and quantum
mechanical probability densities for momentum. We will also discuss a very
useful interpolation between the symmetric linear well, defined by
the potential $V(x) = F|x|$ (and hence analogous to the 'bouncer'), and the 
standard infinite well which can be used to show how the position- and
momentum-space distributions smoothly go from one limit to the other. We
first recall that if we define the infinite potential with walls at
$x = \pm a$, via
\begin{equation}
      V(x) = \left\{ \begin{array}{cl}
               +\infty & \mbox{for $|x|>a$} \\
               0  & \mbox{for $|x|<a$}
                                \end{array}
\right.
\end{equation}
the classical speed and 'half-period' are trivially related to the
total energy via $v = \sqrt{2E/m}$ and $\tau = 2a/v$ so that the
classical probability density for position is given 
\begin{equation}
P_{CL}(x) = \frac{1}{\tau v(x)} = \frac{1}{2a}
\end{equation}
independent of $E$ and determined solely by the geometry of the well.
The energy eigenstates for the corresponding quantum mechanical problem
can be characterized by parity with,  for example, the even states being
given by
\begin{equation}
\psi_n^{(+)}(x) = \frac{1}{\sqrt{a}}\cos\left(\frac{(n-1/2)\pi x}{a}\right)
\qquad
\mbox{with energies}
\qquad
E_n^{(+)} = \frac{ \hbar^2 (n-1/2)^2}{8ma^2}
\end{equation}
and similar expressions for the odd states.
For large values of $n$, the quantum mechanical probability density,
$P_{QM}(x) = |\psi_n^{(+)}(x)|^2$,  locally averages to the classical result, 
since $\langle \cos^2(z) \rangle = 1/2$.

The classical momentum probability density can be easily described in words 
as consisting of equal probabilities of finding
the particle with momentum values given by $\pm p_0 = \pm \sqrt{2mE}$, 
which one can write in the form
\begin{equation}
P_{CL}(p) = \frac{1}{2}\left[\delta(p-p_0) + \delta(p+p_0)\right]
\,.
\end{equation}
The corresponding quantum mechanical solutions mimic this 'twin peaks' 
structure  as one can see by calculating the momentum-space wave function
for the even case, e.g.
\begin{eqnarray}
\phi_n^{(+)}(p) & = & \prefactor \int_{-\infty}^{+\infty} \, \psi_n^{(+)}(x)\,dx 
\nonumber \\
& = & \sqrt{\frac{a}{2\pi \hbar}}
\left[
\frac{\sin((n-1/2)\pi - ap/\hbar)}{((n-1/2)\pi - ap/\hbar)}
+
\frac{\sin((n-1/2)\pi + ap/\hbar)}{((n-1/2)\pi + ap/\hbar)}
\right]
\label{quantum_momentum_infinite_well}
\\ 
& = & 
\sqrt{\frac{\Delta p}{2\pi}}
\left[
\frac{\sin((p_n - p)/\Delta p)}{(p_n-p)}
+
\frac{\sin((p_n + p)/\Delta p)}{(p_n+p)}
\right]
\nonumber
\end{eqnarray}
where $p_n \equiv (n-1/2)\pi\hbar/a$ and $\Delta p \equiv \hbar/a$.
In a classical limit,  where we might take $\hbar \rightarrow 0$ or
 $a \rightarrow \infty$ (or at least much larger than any quantum 
dimension), we have $\Delta p \rightarrow 0$ and one can show that
\begin{equation}
\lim_{\hbar/a \rightarrow 0} |\phi_n^{(+)}(p)|^2 = \frac{1}{2}
\left[\delta(p-p_n) + \delta(p+p_n\right]
\end{equation}
as expected. 

To students first encountering such concepts, however, the use of the
mathematical formalism of the Dirac $\delta$-function can be less than
intuitively obvious, so it would be useful to have an appropriate limiting
case where students can see, both physically and mathematically, how the
infinite well case is reached. To this end, we consider a variation on the
symmetric linear  well, $V(x) = F|x|$, namely one with infinite walls 
included, defined via
\begin{equation}
      V(x) = \left\{ \begin{array}{ll}
               F|x| = V_0|x|/a & \mbox{for $|x|<a$} \\
               +\infty  & \mbox{for $|x|>a$}
                                \end{array}
\right.
\,.
\label{closed_court_potential}
\end{equation}
This potential is sometimes called the 'bouncer on a closed court'
\cite{closed_court}. 
In the limit that $V_0 \rightarrow 0$, this potential approaches the
symmetric infinite well. We will especially be interested in cases where
the energy is greater than $V_0$ (as in Figs.~3, 4, and 5). In that
case, because of the infinite walls, the particle will never slow down and
come to rest as it reverses its direction, but rather simply 'bounce' off
the infinite wall. In this case, there is not only a maximum value of
momentum given by $p_{(+)} = \sqrt{2mE}$, but also a minimum value given by
$p_{(-)} = \sqrt{2m(E-V_0)}$, so that the classical probability density for
momenta will only be non-vanishing over the intervals $(-p_{+},-p_{-})$ and
$(+p_{-},+p_{+})$. The 'half-period', $\tau$,  required for the position 
probability density is given by (so long as $E>V_0$,  as we assume)
\begin{eqnarray}
\tau & = &\int_{-a}^{+a}\frac{dx}{v(x)} 
 =  \int_{-a}^{+a} \frac{dx}{\sqrt{2(E-V_0|x|/a)/m}} \\
& = & 2\sqrt{\frac{m}{2}}\left(\frac{4a}{V_0}\right)
\left[\sqrt{E} - \sqrt{E-V_0}\right] \nonumber
\end{eqnarray}
and the classical period is $T_{CL} = 2\tau$. Since the particle 
achieves the same momentum value twice during each period (due to the
symmetry of the well) and $|F| = V_0/a$, we can write
\begin{equation}
P_{CL}(p) = \frac{2}{T_{CL}|F(p)|}
= \frac{1}{2} \left[\frac{1}{\sqrt{2mE} - \sqrt{2m(E-V_0)}}\right]
= \frac{1}{2} \left[\frac{1}{p_{(+)} - p_{(-)}}\right]
\end{equation}
over the two allowed regions and vanishing elsewhere.
The complete expression for the classical probability density for momentum 
can therefore be written in the form
\begin{equation}
      P_{CL}(p)  = \left\{ \begin{array}{ll}
               \frac{1}{2\Delta p} & \mbox{for $p_{(-)} \leq |p| \leq p_{(+)}$} \\
               0 & \mbox{otherwise}
                                \end{array}
\right.
\label{pclp}
\end{equation}
where $\Delta p \equiv p_{(+)} - p_{(-)}$ and $P_{CL}(p)$ is clearly
normalized appropriately.
For fixed values of $E$ and $a$, as we let $V_0 \rightarrow 0$, we have
$\Delta p \rightarrow 0$ and we clearly reproduce the $\delta$-function 
structure of the infinite well as the two, isolated rectangular peaks become
narrower and higher. 

The classical probability density for position can also be easily
derived and we find that
\begin{equation}
P_{CL}(x) = \frac{1}{\tau v(x)} =
\frac{V_0}{4a[\sqrt{E}-\sqrt{E-V_0}]\sqrt{E-V_0|x|/a}}
\qquad
\mbox{for $|x|\leq a$}
\label{pclx}
\end{equation}
which has the appropriate limit, namely $P_{CL}(x)=1/2a$,
 when $V_0 \rightarrow 0$.

In order to compare these classical predictions to the corresponding quantum
mechanical solutions, we note that the general solutions to the Schr\"odinger 
equation for this problem can be written in the form
\begin{equation}
      \psi(x) = \left\{ \begin{array}{ll}
              C_L Ai(-(x+\sigma)/\rho) + D_L Bi(-(x+\sigma)/\rho)  
& \mbox{for $-a\leq x\leq 0$} \\
C_R Ai((x- \sigma)/\rho) + D_R Bi((x - \sigma)/\rho)  
                & \mbox{for $0\leq x\leq +a$}
                                \end{array}
\right.
\end{equation}
where $Ai(z)$ and $Bi(z)$ are standard Airy functions \cite{handbook}
and where we have defined
\begin{equation}
\rho = \left(\frac{\hbar^2 a}{2mV_0}\right)^{1/3}
\qquad
\mbox{and}
\qquad
\sigma = \frac{Ea}{V_0}
\label{odd_eigenvalue_definitions}
\end{equation}
We note that symmetry arguments can be used to relate the $L$ and $R$ 
coefficients for the even or odd eigenstates. The divergent $Bi(z)$ solutions 
must be used in order to properly match the boundary conditions both 
at the infinite walls and at the origin. Application of the boundary conditions
at $x=0$ and $x=a$, for example, gives the energy eigenvalue condition
for the odd states (where $\psi(0) = 0$) as
\begin{equation}
Ai\left(\frac{-\sigma}{\rho}\right)Bi\left(\frac{a-\sigma}{\rho}\right)
-
Ai\left(\frac{a-\sigma}{\rho}\right)Bi\left(\frac{-\sigma}{\rho}\right) 
= 0
\label{airy_energy_condition}
\end{equation}
and a similar condition for the even eigenstates. The corresponding
momentum eigenstates can then be obtained by direct Fourier transform
from the allowed $\psi(x)$.

As examples of how the limiting case of the infinite well can be approached
from the symmetric linear potential, and of how well the classical and
quantum probability densities agree (correspondence principle limit), 
we consider several cases with
decreasing values of $V_0$. Specifically, in Figs.~3, 4 and 5, we show
examples of solutions (both classical and quantum mechanical) for position-
and momentum-space probability densities using Eqns.~(\ref{pclp}) and
(\ref{pclx}) for the classical versions and the appropriate quantum 
solutions. In each case, we have picked an energy corresponding to an allowed
quantum eigenstate with $E \geqnew 10$ (in scaled units) so that the
maximum value of momentum, $p_{(+)} = \sqrt{2mE}$ is virtually identical in
all three cases. In each case we use $\hbar = 2m = 1$ for simplicity,
and choose $a=25$, while varying the values of $V_0$. 
We then show results for $V_0 = 10$ (Fig.~3), $V_0 = 6$ (Fig.~4) and
$V_0 = 2$ (Fig.~5) for which the appropriate parameters are shown in
Table~I.

We note that as $V_0 \rightarrow 0$, both the classical
and quantum position probability densities, $P_{CL}(x)$ and $P_{QM}(x)$, 
 approach the appropriate 'flat' result for
the infinite well: in each case we have used the same vertical scale to
show this progression. This limiting behavior of  $\psi(x)$ 
as $V_0 \rightarrow 0$ can be 
examined more rigorously using 'handbook' \cite{handbook} properties of the 
$Ai(z)$ and $Bi(z)$ functions.

For the case of the momentum densities, the
approach to the highly peaked case of the infinite well is also clear, with
decreasingly small values of $\Delta p = p_{(+)} - p_{(-)}$ and increasingly
large 'spikes': for these cases, in order to see the detailed structure, we
have {\bf not} shown the momentum densities with the same vertical scales
(note the axis labeling carefully), 
but have kept the horizontal axes identical. Interested students can 
study the
systematic variations and correlations between these three cases and
may especially notice how the spatial variations in the frequency of 
zero-crossings in the 
$|\psi(x)|^2$ are  correlated with the allowed range in momenta.

As a final comment, we note that the 'intrinsic' spread of the two prominent
 individual peaks in the momentum space probability densities given by
Eqn.~(\ref{quantum_momentum_infinite_well}) is of the order of $\Delta p
= \hbar/a$. This means that when the difference $\Delta p = p_{(+)} - p_{(-)}$
becomes of this order, the classical description will clearly be a bad
representation of the quantum results. In the numerical studies 
represented in Figs.~3, 4, and 5, with the specific parameters we have
used for illustrative purposes, we have $\Delta p = \hbar/a = 1/25
= 0.04$ which is still substantially smaller than the 
$\Delta p = p_{(+)} - p_{(-)}$ values  shown in Table~I.

\begin{flushleft}
{\large {\bf 4.~Conclusions}}
\end{flushleft}
\vskip 0.2cm

We have examined the structure of classical probability densities
for momentum, $P_{CL}(p)$, in several simple variations on familiar,
one-dimensional potential problems. 
As our main example, we have shown how a simple 
interpolation between the symmetric linear
potential and the infinite well, provided by 
Eqn.~(\ref{closed_court_potential}) can illustrate, in an intuitively
attractive and mathematically appropriate manner,  how the divergent
$\delta$-function form of the classical probability density for momentum
can arise in the most familiar of all bound state problems, namely the
infinite well. Such examples can hopefully provide experience on how
to understand and visualize the information  content of both classical
and quantum mechanical bound state problems found in a momentum-space
description of these systems, a topic which is seldom covered in detail
in standard undergraduate texts,
 but  which is increasingly important in the
modern research literature.

\begin{flushleft}
{\large {\bf Acknowledgments}}
\end{flushleft}
\vskip 0.2cm

The work of RR was supported, in part, by NSF grant DUE-9950702.

\newpage

\begin{flushleft}
{\Large {\bf 
Tables}}
\end{flushleft}
\vskip 1.0cm

\begin{center}
\begin{tabular}{|c|c|c|c|c|c|} \hline
$V_0$  & $a$  & $E$      & $p_{(-)}$  & $p_{(+)}$ & $\Delta p$ \\ \hline
$10$   & $25$ & $10.066$ & $0.257$    & $3.173$   & $2.916$ \\ \hline
$6$    & $25$ & $10.073$ & $2.108$    & $3.174$   & $1.156$  \\ \hline
$2$    & $25$ & $10.105$ & $2.847$    & $3.179$   & $0.332$ \\ \hline
\end{tabular}
\end{center}

\vskip 1cm
\noindent
{\bf Table~I}.  Values of potential well parameters for the
'closed-court' potential of Eqn.~(\ref{closed_court_potential}) 
used in Figs.~3, 4, and 5 respectively. We use $\hbar = 2m = 1$ in each case. 
The energy eigenvalues are chosen to be as similar as possible,  so that 
the values of $p_{(+)}$ are  virtually
identical in all three cases.

\newpage

\begin{flushleft}
{\Large {\bf 
Figure Captions}}
\end{flushleft}
\vskip 0.5cm
 
\begin{itemize}
\item[Fig.\thinspace 1.]  Plot of a generic potential energy function,
$V(x)$ versus $x$,  which supports bound state motion. A constant value
of the total energy, $E$, defines the classical turning points $a,b$. The
corresponding phase-space $x-p$ plot for the system is shown directly
below. The extremal values of momentum, $\pm p_m = \pm \sqrt{E-V_{min}}$, 
correspond to the  position of the minimum of the potential energy function.  
The $x-p$ phase-space plot is automatically symmetric about the $p=0$ axis 
due to the equal probabilities of finding the particle moving through the
same $dx$ bin,  to the left or right,
 at two different times during its classical period.
\item[Fig.\thinspace 2.] Illustration of the 'projection of trajectory'
technique for visualizing the classical probability densities for position
(top) and momentum (bottom) for the classical 'bouncer'. Also shown are
$N=1000$ measurements of the particles position ($z$) and momentum ($p_z$)
coordinate,  which are consistent with the binned probabilities shown on the
vertical $z(t)$ and $p_z(t)$ axes.
\item[Fig.\thinspace 3.] Potential energy function (top), 
classical ($P_{CL}(x)$, dashed) and quantum ($P_{QM}(x) = |\psi(x)|^2$, solid)
position probability densities  (middle), and classical ($P_{CL}(p)$, dashed) 
and quantum ($P_{QM}(p) = |\phi(p)|^2$, solid) momentum probability 
densities (bottom) for the symmetric linear well with infinite wall potential 
in Eqn.~(\ref{closed_court_potential}) with $V_0 = 10$. The values of the
other parameters are shown in Table~I.
\item[Fig.\thinspace 4.] Same as Fig.~3, but with $V_0 = 6$. Note that
the vertical axis for the $P_{CL}(p), P_{QM}(p)$ plot is not the
same as in Figs.~3 or 5.
\item[Fig.\thinspace 5.] Same as Fig.~3, but with $V_0 = 2$. Note that
the vertical axis for the $P_{CL}(p), P_{QM}(p)$ plot is not the
same as in Figs.~4 or 5.
\end{itemize}

\newpage

\noindent
\hfill
\begin{figure}[hbt]
\, \hfill \,
\begin{minipage}{0.7\linewidth}
\epsfig{file=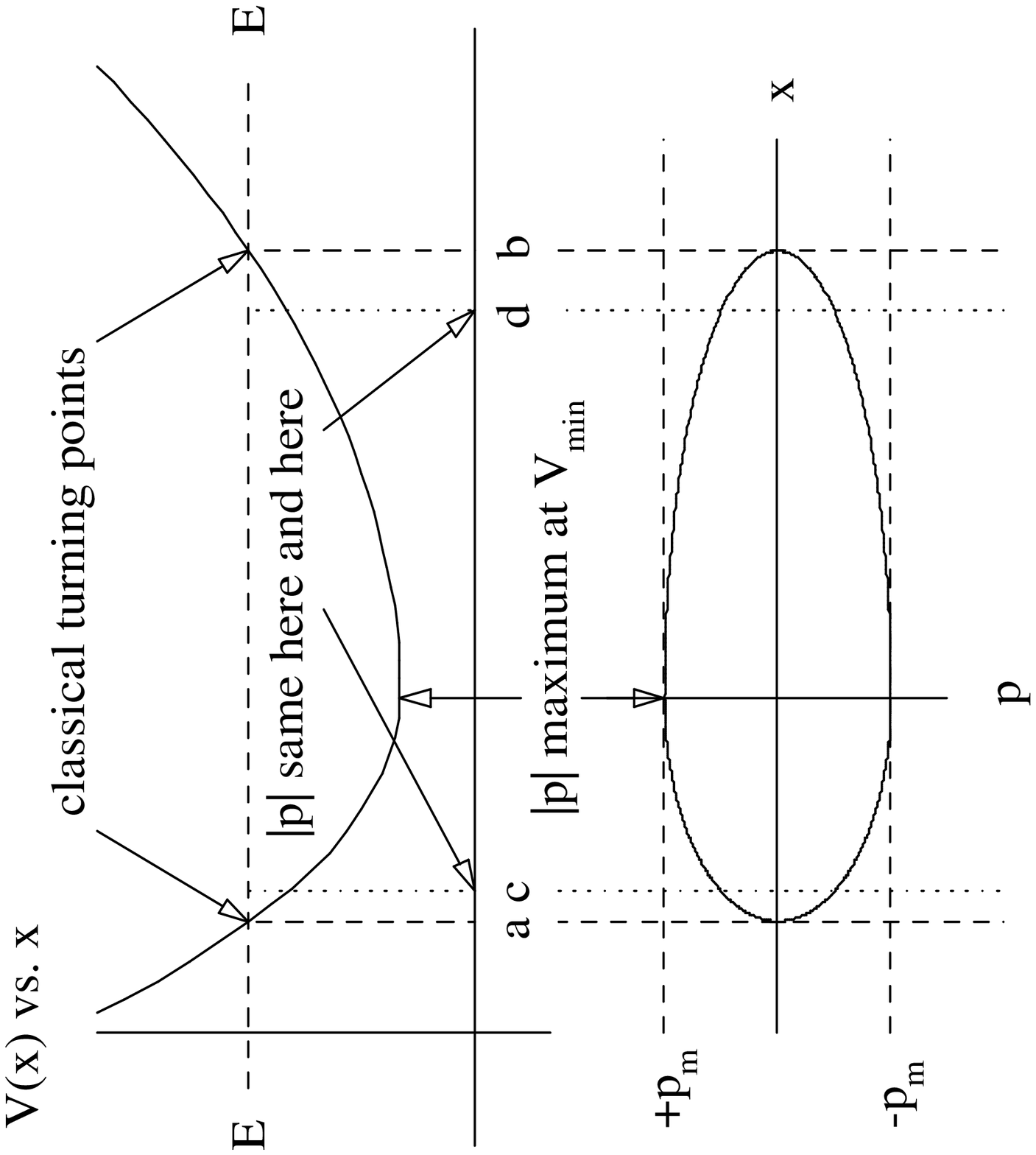,width=\linewidth}
\caption{}
\end{minipage}
\, \hfill \,
\end{figure}
\hfill

\newpage

\noindent
\hfill
\begin{figure}[hbt]
\, \hfill \,
\begin{minipage}{0.7\linewidth}
\epsfig{file=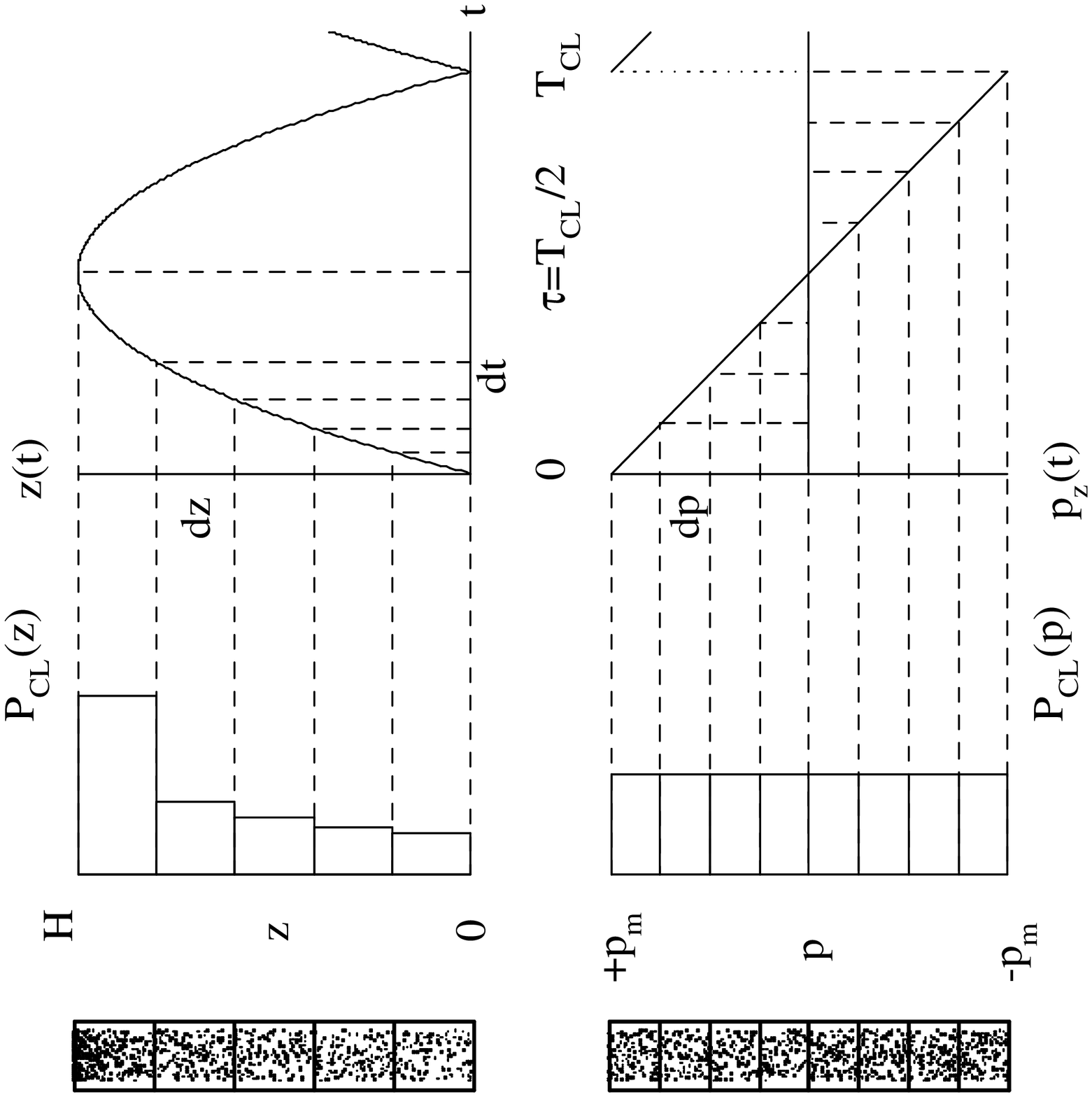,width=\linewidth}
\caption{}
\end{minipage}
\, \hfill \,
\end{figure}
\hfill

\newpage

\noindent
\hfill
\begin{figure}[hbt]
\, \hfill \,
\begin{minipage}{0.7\linewidth}
\epsfig{file=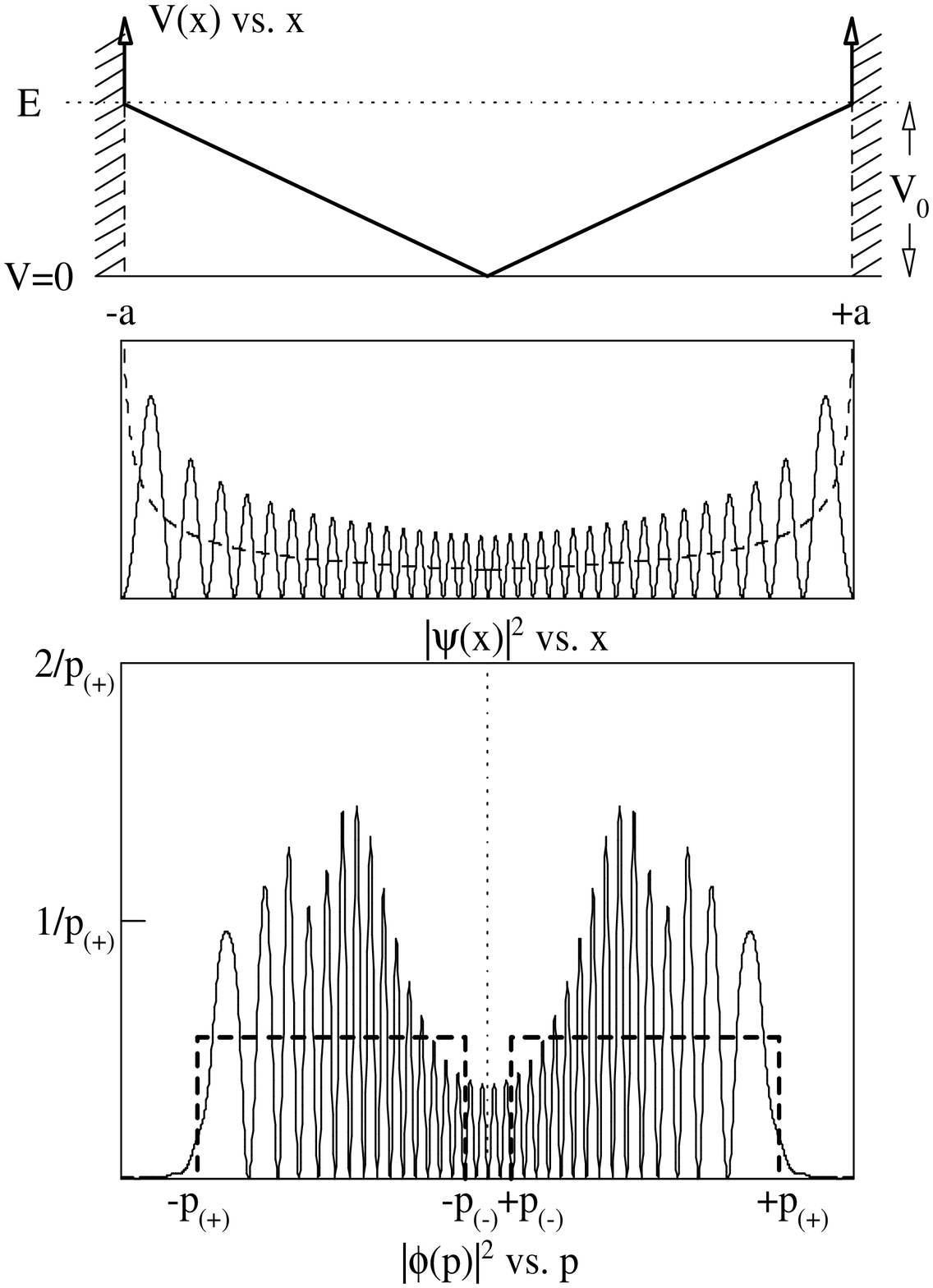,width=\linewidth}
\caption{}
\end{minipage}
\, \hfill \,
\end{figure}
\hfill

\newpage

\noindent
\hfill
\begin{figure}[hbt]
\, \hfill \,
\begin{minipage}{0.7\linewidth}
\epsfig{file=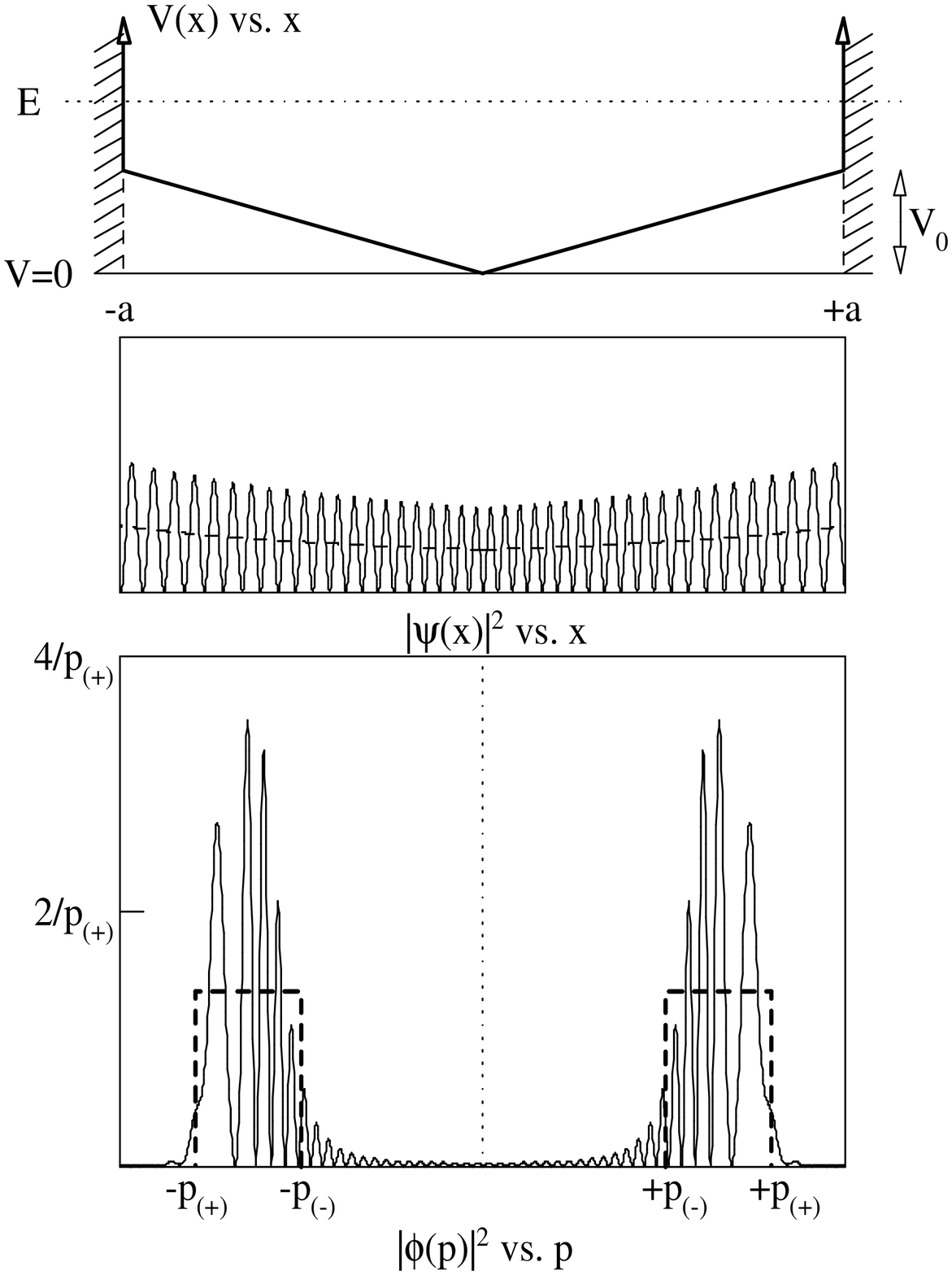,width=\linewidth}
\caption{}
\end{minipage}
\, \hfill \,
\end{figure}
\hfill

\newpage

\noindent
\hfill
\begin{figure}[hbt]
\, \hfill \,
\begin{minipage}{0.7\linewidth}
\epsfig{file=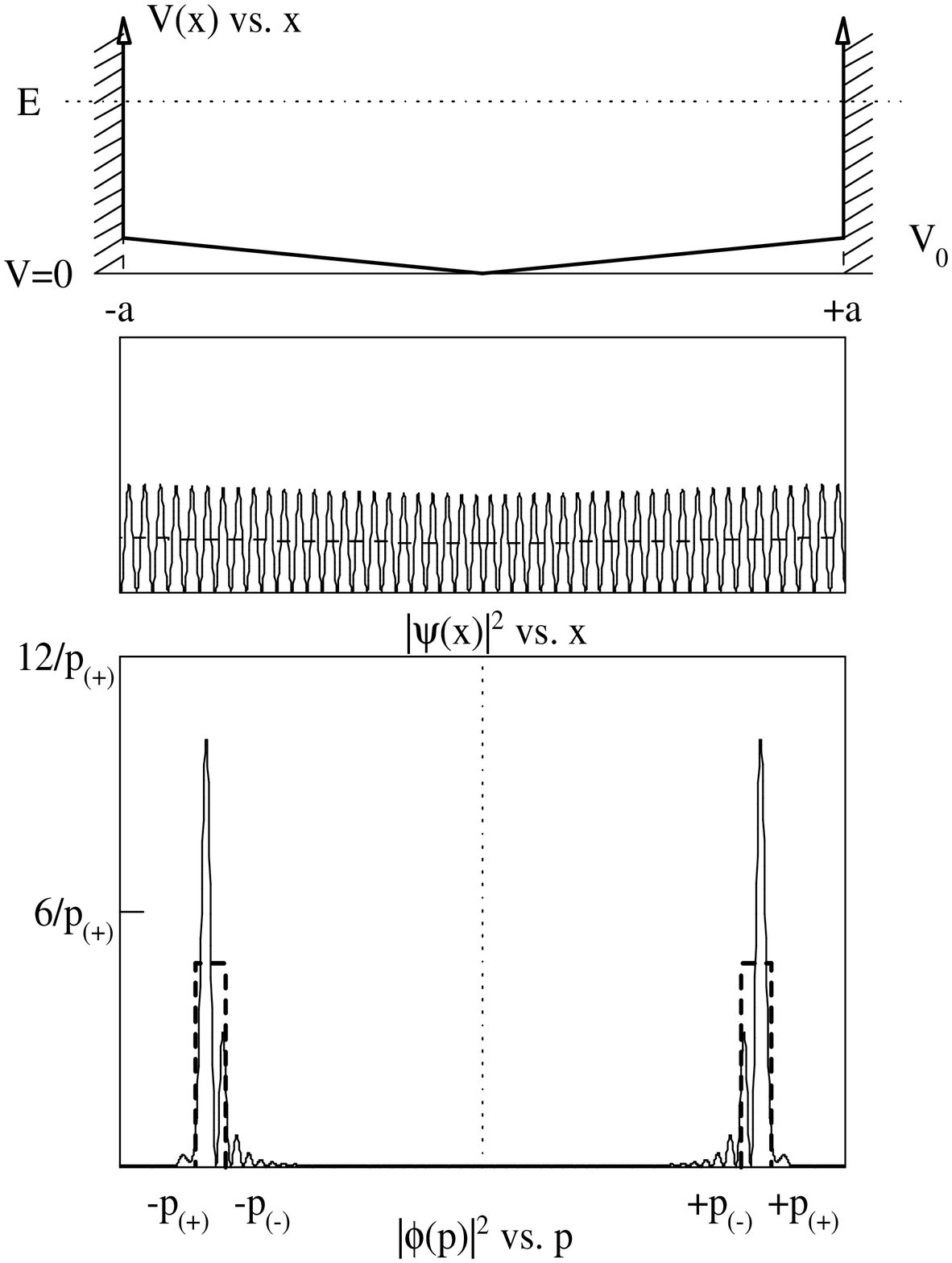,width=\linewidth}
\caption{}
\end{minipage}
\, \hfill \,
\end{figure}
\hfill


\newpage
 
\begin{thebibliography}{99}
%
%
\bibitem{park} See, e.g., Park~D 1992 {\it Introduction to the Quantum
Theory} (New York: McGraw-Hill) 3rd edition pp~239-241
%
%
\bibitem{liboff} Liboff~R~L 1997 {\it Introductory Quantum Mechanics}
(Reading: Addison-Wesley) 3rd edition pp~192-195
%
%
\bibitem{robinett_book} Robinett~R~W 1997 {\it Quantum Mechanics: Classical
Results, Modern Systems, and Visualized Examples} (New York: Oxford University
Press)  pp~110-115
%
%
\bibitem{qmvi} Cataloglu E and Robinett~R~W 2001 Testing the development
of student conceptual and visualization understanding in quantum mechanics
through the undergraduate career (to appear in American Journal of Physics,
theme issue on quantum mechanics).
%
%
\bibitem{modern_1} Hensinger~W~K {\it et al.} 2001 Dynamical tunneling of
ultracold atoms Nature {\bf 412} 52-55
%
%
\bibitem{modern_2} Steck~A~S {\it et al.} 2001 Observation of chaos-assisted
tunneling between islands of stability Science {\bf 293} 274-278
%
%
\bibitem{bouncer_6} Many references to the 'bouncer' problem in the
pedagogical literature can be found in Gea-Banacloche~J 1999
A quantum bouncing ball Am. J. Phys.  {\bf 67} pp~410-420
%
%
\bibitem{robinett_ajp} Robinett~R~W 1995  Quantum and classical probability
distributions for position and momentum  Am. J. Phys. {\bf 63} 823-832 
%
%
\bibitem{closed_court} Aguilera-Navarro~V~C, Iwamoto~H, Ley-Koo~E,
and Zimerman~A~H 1981 Quantum bouncer in a closed court Am. J. Phys.
{\bf 49} pp~648-651
%
%
\bibitem{handbook} Abramowitz M and Stegun I A 1964
Handbook of Mathematical Functions
(New York: McGraw-Hill) pp.~446-449
%
%
\end{thebibliography}
\end{document}